\documentclass[traditabstract]{aa} 
\usepackage{amsmath}
\usepackage{graphicx}
\usepackage{txfonts}
\usepackage{natbib,twoopt}
\usepackage[breaklinks=true]{hyperref} 
\bibpunct{(}{)}{;}{a}{}{,} 
\makeatletter
\newcommandtwoopt{\citeads}[3][][]{\href{http://adsabs.harvard.edu/abs/#3}%
{\def\hyper@linkstart##1##2{}%
\let\hyper@linkend\@empty\citealp[#1][#2]{#3}}}
\newcommandtwoopt{\citepads}[3][][]{\href{http://adsabs.harvard.edu/abs/#3}%
{\def\hyper@linkstart##1##2{}%
\let\hyper@linkend\@empty\citep[#1][#2]{#3}}}
\newcommandtwoopt{\citetads}[3][][]{\href{http://adsabs.harvard.edu/abs/#3}%
{\def\hyper@linkstart##1##2{}%
\let\hyper@linkend\@empty\citet[#1][#2]{#3}}}
\newcommandtwoopt{\citeyearads}[3][][]%
{\href{http://adsabs.harvard.edu/abs/#3}
{\def\hyper@linkstart##1##2{}%
\let\hyper@linkend\@empty\citeyear[#1][#2]{#3}}}
\makeatother
\begin{document}

   \title{Anomalous orbital expansion of low-mass X-ray binary 2A 1822-371: the existence of a circumbinary disk?}

   \titlerunning{Anomalous orbital expansion of low-mass X-ray binary 2A 1822-371}

   \authorrunning{N. Wei, L. Jiang \& W.-C. Chen }

  \author{Na Wei$^{1}$, Long Jiang$^{1,2}$, and Wen-Cong Chen$^{1,2}$}

  \institute{$^{1}$School of Science, Qingdao University of Technology, Qingdao 266525, China\\
 $^{2}$School of Physics and Electrical Information, Shangqiu Normal University, Shangqiu 476000, China}

   \date{Received 14 March 2023 / Accepted * * 2023}

\abstract{The source 2A 1822-371 is an eclipsing low-mass X-ray binary (LMXB) consisting of a neutron star (NS) and a $\sim0.5~M_{\odot}$ donor star in an orbit of 5.57 hr. Based on timing of the eclipse arrival times, this source was found to be experiencing a rapid orbital expansion with an orbital-period derivative as $\dot{P}_{\rm orb}=(1.51\pm0.05)\times10^{-10}~\rm s\, s^{-1}$, implying that the mass-transfer rate should be higher than at least three times the Eddington accretion rate. The standard magnetic braking (MB) model cannot produce such a high mass-transfer rate. The modified MB model derived by Van \& Ivanova (2019) can produce a high mass-transfer rate, resulting in a high $\dot{P}_{\rm orb}$. This work proposes an alternative model to account for the anomalously high mass-transfer rate and $\dot{P}_{\rm orb}$ of 2A 1822-371. During the mass transfer, a tiny fraction of the transferred material is thought to form a circumbinary (CB) disk around the LMXB, which can efficiently extract orbital angular momentum from the system by the interaction between the CB disk and the binary. We use the MESA code to model the formation and evolution of 2A 1822-371 for different CB-disk masses. When the CB-disk mass is $2.3\times10^{-8}~ M_{\odot}$, the simulation can reproduce the observed donor-star mass, orbital period, and orbital-period derivative. Such a CB disk can accelerate the evolution of the binary and produce a high mass transfer rate of $1.9\times10^{-7}~ M_\odot\,\rm yr^{-1}$, driving the binary to evolve toward a wide-orbit system. Therefore, we propose that CB disks may be responsible for the rapid orbital changes observed in some LMXBs.}

\maketitle
\section{Introduction}
According to the masses of the companion stars, strong Galactic X-ray sources are mainly classified into two groups: high-mass
X-ray binaries and low-mass X-ray binaries (LMXBs) \citep{taur06}. In LMXBs, a neutron star (NS) or black hole accretes materials from its donor star with a mass less than $1-2~M_{\odot}$ by Roche lobe overflow (RLOF) and emits strong X-ray. Due to the long accretion phase in LMXBs, the accreting NSs are spun up to the millisecond period, and evolve into millisecond pulsars \citep{alpa82,radh82}. Therefore, studying LMXBs is significant in understanding the formation and evolution of millisecond pulsars. Furthermore, LMXBs provide much information about mass exchange and angular momentum loss, thus they are ideal probes testing stellar and binary evolution theory.

The source 2A 1822-371 (sometimes named X 1822-371) is a persistent eclipsing LMXB with a short orbital period of 5.57 hr \citep{maso80,whit81}, in which a NS with a relatively small spin frequency of $\nu\sim1.69~\rm Hz$ \citep{jonk01} is spinning up at a rate of $\dot{\nu}=(7.57\pm0.06)\times10^{-12}~\rm Hz\, s^{-1}$ via a mass accretion \citep{jain10,iari15,bak17}. According to the partial eclipses, 2A 1822-371 was hinted to be an accretion disk corona source \citep{whit82}, and the inclination angle was inferred to be $81^{\circ}-85^{\circ}$ \citep{hein01,ji11}.

The mass function of 2A 1822-371 was measured to be $f(M)=(2.03\pm0.03)\times10^{-2}~ M_{\rm \odot}$ \citep{jonk01}, resulting in a minimum companion-star mass of $M_{\rm d}=0.33\pm0.05~ M_{\rm \odot}$ \citep{jonk03}. By studying the K-correction for the emission lines formed in the X-ray-illuminated atmosphere of the donor star, the masses of the NS and the donor star were constrained to be
$1.61~ M_{\rm \odot}\leq M_{\rm NS}\leq 2.32~ M_{\rm \odot}$ and $0.44~ M_{\rm \odot}\leq M_{\rm d}\leq 0.56~ M_{\rm \odot}$, respectively \citep{muno05}.

According to a distance of 2.5 kpc, the unabsorbed X-ray luminosity of 2A 1822-371 was calculated to be $L_{\rm X}\simeq 1.0\times10^{36}~\rm erg\,s^{-1}$, yielding a mean ratio of the X-ray over optical luminosity as $L_{\rm X}/L_{\rm opt}\sim20$ \citep{maso82}. Such a ratio is much smaller than the typical luminosity ratio ($L_{\rm X}/L_{\rm opt}\sim1000$) of LMXBs, implying that the intrinsic luminosity of 2A 1822-371 may exceed the Eddington limit \citep{bayl10,burd10}. Magnetic braking (MB) is the main mechanism driving the mass transfer of LMXBs with long orbital periods. However, the standard MB model given by \cite{rapp83} is challenging to produce a super-Eddington mass transfer in this LMXB with a low-mass donor star.

The orbital evolution of the source 2A 1822-371 also remains mysterious. Based on the analysis of eclipses detected by the HEAO-1, Einstein, Exosat, and Ginga, its orbital period was thought to be increasing at a rate of $\dot{P}_{\rm orb}=(2.19\pm0.58)\times10^{-10}~\rm s\, s^{-1}$ \citep{hell90}. According to an improved ephemeris for the
optical eclipses, \cite{bayl10} confirmed that its orbital period is rapidly changing at a rate of $\dot{P}_{\rm orb}=2.12\times10^{-10}~\rm s\, s^{-1}$. Subsequently, the orbital period derivative of 2A 1822-371 was independently measured to be $\dot{P}_{\rm orb}=(1.50\pm0.07)\times10^{-10}~\rm s\, s^{-1}$ \citep{burd10}, $\dot{P}_{\rm orb}=(1.59\pm0.09)\times10^{-10}~\rm s\, s^{-1}$ \citep{iari11}, $\dot{P}_{\rm orb}=(1.464\pm0.041)\times10^{-10}~\rm s\, s^{-1}$ \citep{chou16}, and $\dot{P}_{\rm orb}=(1.475\pm0.054)\times10^{-10}~\rm s\, s^{-1}$ \citep{mazz19}. Recently, the orbital period derivative was refined to be $\dot{P}_{\rm orb}=(1.51\pm0.05)\times10^{-10}~\rm s\, s^{-1}$ based on the updated orbital ephemeris joining two new eclipse times related to NuSTAR and Swift observations \citep{anit21}. Such an orbital-period derivative is three orders of magnitude higher than that derived from conservative mass transfer driven by MB and gravitational radiation, implying that the mass-transfer rate should be higher than at least three times the Eddington accretion rate of a NS \citep{burd10,bayl10}. As mentioned above, the standard MB model cannot produce such a high mass-transfer rate for a low-mass donor star. Therefore, the formation and evolution of the source 2A 1822-371 challenge the conventional binary evolution theory, which may require a new MB description or new physical mechanisms.

\section{Analysis for the Orbital Evolution}
The total orbital angular momentum of a LMXB consisting of a NS and a low-mass donor star can be written as
\begin{equation}
J=\frac{M_{\rm NS}M_{\rm d}}{M_{\rm NS}+M_{\rm d}}\Omega{a^{2}},
\end{equation}
where $M_{\rm NS}$ and $M_{\rm d}$ are the NS mass and the donor-star mass, respectively; $\Omega=2\pi/P_{\rm orb}$ is the orbital angular velocity of the binary, and $a$ is the orbital separation. Inserting Kepler's third law ($G(M_{\rm NS}+M_{\rm d})/a^{3}=\Omega^{2}$) into equation (1) and differentiating it, the orbital period derivative satisfies
\begin{equation}
\frac{\dot{P}_{\rm orb}}{P_{\rm orb}}=3\frac{\dot{J}}{J}-3\frac{\dot{M}_{\rm d}}{M_{\rm d}}\left(1-q\beta-\frac{q(1-\beta)}{3(1+q)}\right),
\end{equation}
where $\dot{x}={\rm d}x/{\rm d}t$, $q=M_{\rm d}/M_{\rm NS}$ is the mass ratio of the binary, and $\beta=-\dot{M}_{\rm NS}/\dot{M}_{\rm d}$ is the accretion efficiency of the NS.
\subsection{conservative mass transfer}
In the case of conservative mass transfer, $\beta=1$, equation (2) changes into
\begin{equation}
\frac{\dot{P}_{\rm orb}}{P_{\rm orb}}=3\frac{\dot{J}}{J}-3\frac{\dot{M}_{\rm d}}{M_{\rm d}}(1-q).
\end{equation}
The first term on the right-hand side of equation (3) would produce a negative $\dot{P}_{\rm orb}$ because $\dot{J}<0$, while
the mass transfer from the less massive donor star to the more massive NS would cause a positive $\dot{P}_{\rm orb}$ since $\dot{M}_{\rm d}<0$ and $q<1$ (see also the second term on the right-hand side). The angular-momentum-loss rate predicted by the standard MB model is
\begin{equation}
\dot{J}_{\rm mb}=-3.8\times10^{-30}M_{\rm d}R_{\odot}^{4}(R_{\rm d}/R_{\odot})^\gamma\Omega^3~\rm dyn\,cm,
\end{equation}
where $R_{\rm d}$ is the donor-star radius, $\gamma$ is a dimensionless magnetic-braking index from 0 to 4 \citep{rapp83}. In this work, we adopt the simplest approximation as $\gamma=4$ \citep{verb81}. Since the donor star filled its Roche lobe, its radius can be estimated to be \citep{eggl83}
\begin{equation}
R_{\rm d}\approx R_{\rm L}=\frac{0.49q^{2/3}}{0.6q^{2/3}+ {\rm
ln}(1+q^{1/3})}a.
\end{equation}

For a binary with an orbital period exceeding 3 hr, the angular-momentum-loss rate by gravitational radiation is weaker than that by MB. Hence the contribution of gravitational radiation can be ignored. Taking $P_{\rm orb}=5.57~\rm hr$, $\dot{P}_{\rm orb}=1.51\times10^{-10}~\rm s\, s^{-1}$, $M_{\rm NS}=1.6~M_{\odot}$, and $M_{\rm d}=0.5~M_{\odot}$, we have $R_{\rm d}=0.57~R_{\odot}$, $a=2.0~R_{\odot}$, $\dot{J}_{\rm mb}/J=-6.3\times10^{-17}~\rm s^{-1}$, and $\dot{P}_{\rm orb}/P=7.5\times10^{-15}~\rm s^{-1}$. Because $|\dot{J}_{\rm mb}/J|$ is two orders of magnitude smaller than $\dot{P}_{\rm orb}/P$, the contribution of angular momentum loss on $\dot{P}_{\rm orb}$ of 2A 1822-371 is trivial. Therefore, according to equation (3), the mass transfer rate can be derived to be $|\dot{M}_{\rm d}|=(5.76\pm0.19)\times10^{-8}~ M_{\odot}\,\rm yr^{-1}$.

\subsection{non-conservative mass transfer}
The calculated mass-transfer rate of 2A 1822-371 is higher than the Eddington accretion rate ($\dot{M}_{\rm Edd}=1.5\times10^{-8}~ M_{\odot}\,\rm yr^{-1}$), hence the mass transfer of 2A 1822-371 is probably nonconservative, i.e. the accretion efficiency $\beta\neq 1$. To study the influence of the accretion efficiency, we plot the relation between the mass transfer rate and the accretion efficiency $\beta$ in Figure 1 according to equation (2) (the contribution of angular momentum loss for $\dot{P}_{\rm orb}$ is still ignored \footnote{In principle, the mass-loss rate exceeding $\dot{M}_{\rm edd}$ would carry away the specific angular momentum of the NS, which contributes $(\dot{J}/J)_{\rm ml}=(1-\beta)\dot{M}_{\rm d}q/[(1+q)M_{\rm ns}]$. Therefore, the ratio between the first term and the second term in the right-hand side of equation (2) is $(1-\beta)q^{2}/[1-q\beta-q(1-\beta)/(3+3q)]/(1+q)$. For 2A 1822-371, this ratio is about 0.075 if we take $q=0.3$ and $\beta=0$; hence $(\dot{J}/J)_{\rm ml}$ can be ignored.}). It is clear that a low accretion efficiency tends to require a low mass-transfer rate to produce the observed orbital-period derivative. For $M_{\rm NS}=1.6~M_{\odot}$ and $M_{\rm d}=0.5~M_{\odot}$, $\beta=1$ predicts a mass-transfer rate of $\approx5.8\times10^{-8}~\rm M_{\odot}\,yr^{-1}$, which is consistent with the mass-transfer rate predicted by the conservative mass transfer case. It seems that the mass-transfer rate is weakly dependent on the accretion efficiency. Even if the NS does not accrete any material ($\beta=0$), it still requires a mass-transfer rate of $\sim(4-5)\times10^{-8}~ M_{\odot}\,\rm yr^{-1}$ to produce the observed $\dot{P}_{\rm orb}$ of 2A 1822-371. Furthermore, the NS mass and the donor-star mass can also alter the derived mass-transfer rate. However, the effect of the latter is more significant than that of the former. Therefore, such a rapid orbital expansion implies a rapid mass transfer between the donor star and the accreting NS of 2A 1822-371.

\begin{figure}
	\centering
	\includegraphics[scale=0.35,trim={0 0 0 0},angle=0]{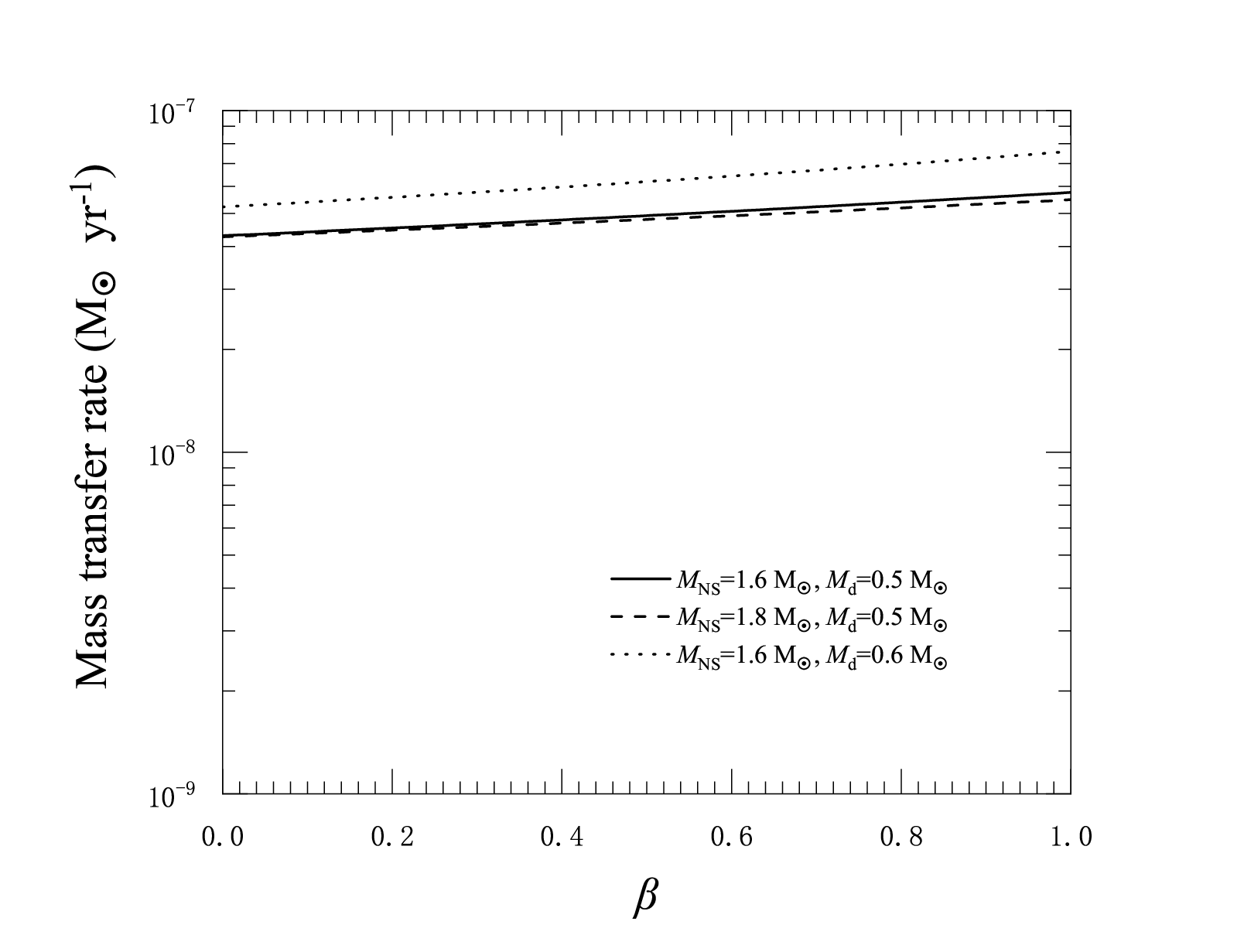}
	\caption{Mass-transfer rate versus accretion efficiency of the NS of 2A 1822-371. We take $P_{\rm orb}=5.57~\rm hr$, $\dot{P}_{\rm orb}=1.51\times10^{-10}~\rm s\, s^{-1}$. The solid, dashed, and dotted curves represent the cases when $(M_{\rm NS},M_{\rm d})=(1.6,0.5),(1.8,0.5)$, and $(1.6,0.6)~M_{\odot}$, respectively. }
	\label{fig:sumAH}
\end{figure}

\subsection{some possible models}
When $q=0.5/1.6\approx0.31$, the standard MB model predicts an approximate mass-transfer rate as \citep{pavl16}
\begin{equation}
  \begin{aligned}
\dot{M}_{\rm d}=-2.0\times10^{-9}\left(\frac{R_{\rm d}}{0.57R_{\odot}}\right)^{4}\left(\frac{2.0R_{\odot}}{a}\right)^2 \\
 \left(\frac{M_{\rm NS}+M_{\rm d}}{2.1M_{\odot}}\right)\left(\frac{1.6M_{\odot}}{M_{\rm NS}}\right)\left(\frac{M_{\rm d}}{0.5M_{\odot}}\right)~ M_{\odot}\,\rm yr^{-1}.
   \end{aligned}
\end{equation}
For some typical parameters of 2A 1822-371, the estimated mass-transfer rate is $2.0\times10^{-9}~ M_{\odot}\,\rm yr^{-1}$, which is about one order of magnitude smaller than that derived from the observed orbital period derivative. Similarly, the inferred mass-transfer rates of some LMXBs with short orbital periods are found to be at least an order of magnitude higher than theoretically expected values \citep{pods02}. To solve this discrepancy problem between the observation and the theory, \cite{van19a} modified MB prescription and proposed a convection and rotation boosted (CARB) MB model (see also section 2.4).

\cite{chou16} argued that the unusual $\dot{P}_{\rm orb}$ may be caused by the X-ray radiation-driven mass loss proposed by \cite{tava93}. Assuming that a fraction ($f$, i.e., the efficiency of irradiation-driving wind) of the X-ray luminosity that the donor star receives conquers the gravitational bind energy of the material on the surface of the donor star and drives a strong wind (with a velocity equaling to the escape
speed of the donor's surface), the loss rate of the irradiation-driving winds is \citep{chen16}
\begin{equation}
  \begin{aligned}
\dot{M}_{\rm wind}= fL_{\rm X}\frac{R_{\rm d}^{3}}{4GM_{\rm d}a^{2}}=1.9\times10^{-8}f_{0.001}L_{\rm X,38}\\
\left(\frac{R_{\rm d}}{0.57R_{\odot}}\right)^{3}
\left(\frac{2.0R_{\odot}}{a}\right)^2\left(\frac{0.5M_{\odot}}{M_{\rm d}}\right)~\rm M_{\odot}yr^{-1},
   \end{aligned}
\end{equation}
where $f_{0.001}=f/0.001$, $L_{\rm X,38}=L_{\rm X}/10^{38}~\rm erg\,s^{-1}$. \cite{tava93} obtained the efficiencies of irradiation-driving wind in the range from 0.001 to 0.1. Therefore, it is possible to produce a relatively high wind-loss rate from the donor star because the intrinsic X-ray luminosity of 2A 1822-371 may exceed $10^{38}~\rm erg\,s^{-1}$.The irradiation-driving winds may also play a key role in resulting in a positive orbital period of the first-discovered accreting millisecond pulsar SAX J1808.4-3658 \citep{chen17}.

Considering angular momentum loss due to MB by coupling between the strong magnetic field and an irradiation-driving wind \citep{just06}, \cite{xing19} used the MESA code to model the evolution of a LMXB consisting of a $1.4~M_{\odot}$ NS and a $ 1.1~M_{\odot}$ donor star in an initial orbit of 0.4 days. Taking a wind-driving efficiency of $f=5\times10^{-3}$ and a magnetic field of $900~\rm G$, their simulated total mass loss rate of the donor star is $\sim10^{-7}~\rm M_{\odot}\,yr^{-1}$ when the donor-star mass is $\sim0.5~M_{\odot}$. Such an irradiation-driving wind model can account for the donor mass, orbital period, orbital-period derivative, and high X-ray luminosity of 2A 1822-371.

\subsection{the CARB MB model}
The CARB MB model considered the influence of the donor-star rotation on the stellar-wind velocity \citep{matt12,revi15} and the influences of the donor-star convective-turnover timescale and the donor-star rotation on its surface magnetic field \citep{park71,noye84,ivan06,van19b}. Included these mechanisms, the angular-momentum-loss rate can be written as \citep{van19a}
\begin{equation}
  \begin{aligned}
\dot{J}_{\rm mb}= -\frac{2}{3}\dot{M}_{\rm w}^{-1/3}R_{\rm d}^{14/3}(v_{\rm esc}^{2}+2\Omega^{2}R_{\rm d}^{2}/K_{2}^{2})^{-2/3}\\
\times\Omega_{\odot}B_{\odot}^{8/3}\left(\frac{\Omega}{\Omega_{\odot}}\right)^{11/3}\left(\frac{\tau_{\rm conv}}{\tau_{\odot,\rm conv}}\right)^{8/3},
   \end{aligned}
\end{equation}
where $\dot{M}_{\rm w}$, and $v_{\rm esc}$ are the wind mass-loss rate, and the surface escape velocity of the donor star, respectively; $K_{2}=0.07$ is a constant originated from a grid of simulations \citep{revi15}; $\tau_{\rm conv}$ is the turnover time of convective eddies; $B_{\odot}=1~\rm G$ is the surface magnetic-field strength of the Sun, and the surface-rotation rate and the convective-turnover time of the Sun are $\Omega_{\odot}\approx3.0\times10^{-6}~\rm s^{-1}$, and $\tau_{\odot,\rm conv}=2.8\times10^{6}~\rm s$ \citep[see also][]{van19b}, respectively. Some information regarding the inlists, and subroutines used to simulate the mass-transfer process of LMXBs in the CARB MB model see also \cite{mang22}.

The CARB MB model can reproduce the observed mass-transfer rates and orbital periods at the detected mass ratio for all observed persistent NS LMXBs in the Galaxy \citep{van19a}. Using the CARB MB model, \cite{van21} investigated the potential progenitors of the observed persistent NS LMXBs, and found that these progenitors occupy a small part of the plausible parameter space in the initial donor-star mass versus initial orbital period diagram. \cite{deng21} also found that the CARB MB model can successfully reproduce the observed characteristics of all persistent NS LMXBs and binary pulsars. Since the CARB MB model can result in a high mass-transfer rate ($\sim10^{-7}~ M_\odot\,\rm yr^{-1}$) for persistent NS LMXBs \citep{van19a,van21}, which is higher than the required mass-transfer rate (see also section 2.2), it is possible to reproduce the observed parameters of 2A 1822-371.

In this paper, we attempt to explore an alternative mechanism that can result in a rapid mass transfer and produce the rapid orbital expansion observed in the source 2A 1822-371. Based on a circumbinary (CB) disk model, we present a detailed stellar evolution model for the formation of 2A 1822-371 in Section 3. In section 4, we discuss the possible influence of input parameters and the observed confirmation. Finally, we give a summary in section 5.

\begin{figure}
	\centering
	\includegraphics[scale=0.34,trim={20 30 0 0},angle=0]{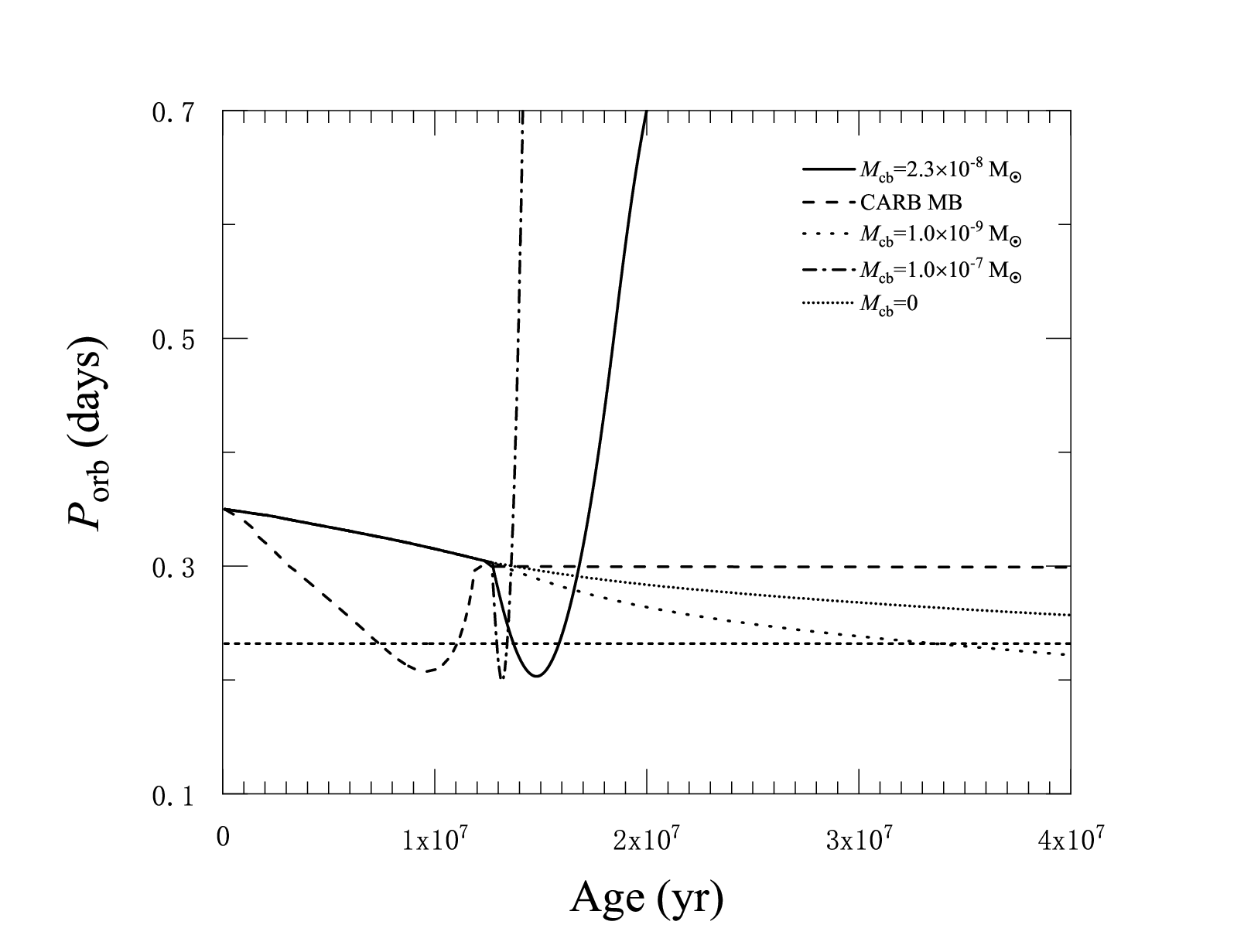}
	\quad
	\includegraphics[scale=0.34,trim={20 30 0 0},angle=0]{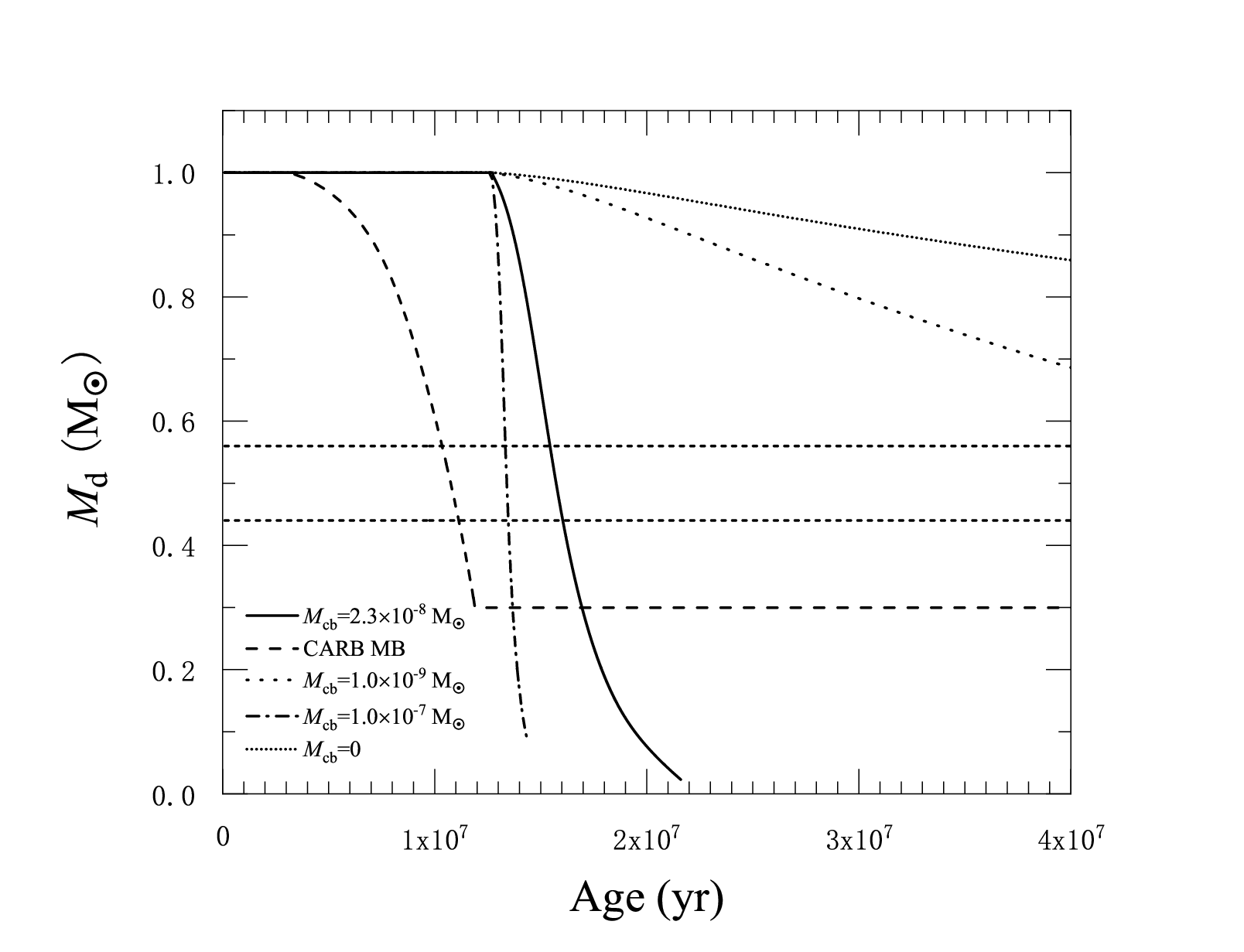}
	\caption{Evolution of a LMXB consisting of $1.6~M_{\rm \odot}$ NS and $1.0~M_{\rm \odot}$ MS donor star with an initial orbit of 0.35 days in the orbital period vs. stellar age diagram (top panel), and the donor-star mass vs. stellar age diagram (bottom panel). The horizontal short dashed line in the top panel indicates the orbital period of 2A 1822-371, while the horizontal short dashed lines in the bottom panel represent the inferred upper limit and lower limit of the donor-star mass.}
	\label{fig:sumAH}
\end{figure}

\section{CB disk Model }
\subsection{CB disk model}

During the mass transfer of a LMXB, a tiny fraction of the transferred material may be ejected by the strong radiation pressure of the accreting NS. Since the ejecta possesses a large orbital angular momentum, they may form a CB disk surrounding the system rather than completely leave it \citep{heuv73,heuv94}.
The resonant torque produced by the interaction between the CB disk and the binary can extract orbital angular momentum from orbital motion. Based on an assumption of a standard thin disk, the angular-momentum-loss rate via a CB disk can be expressed as \citep{chen19}
\begin{equation}
	\dot{J}_{\rm cb}=-M_{\rm cb}\alpha\left(\frac{H}{R}\right)^2\frac{a^3}{R}\Omega^2,
\end{equation}
where $M_{\rm cb}$, $\alpha$, $H$, and $R$ are the mass, viscosity parameter,  thickness, and half angular-momentum radius of the CB disk, respectively.

For simplicity, a CB disk with a constant mass is assumed to surround the LMXB if a mass transfer occurs. Assuming that the ratio between half angular-momentum radius and the orbital separation is a constant ($R/a=2.3$) in three black-hole LMXBs, \cite{chen19} found that the derived CB-disk masses around XTE J1118 and A0620-00 are consistent with the inferred values \citep[$M_{\rm cb}\sim10^{-9}~ M_{\odot}$,][]{muno06} when $\alpha=0.1$, $H/R=0.1$. Same to black-hole LMXBs, we also take $\alpha=0.1$, $H/R=0.1$, and $R/a=2.3$ for 2A 1822-371. CB-disk parameters, including $\alpha$, $H/R$, and $R/a$, might be different for NS LMXBs and black hole LMXBs. However, both $M_{\rm cb}$ and $\alpha(H/R)^{2}R/a$ are degenerate in equation (9). Therefore, some uncertainties resulting from $\alpha$, $H/R$, and $R/a$ can be compensated by slightly altering the CB-disk mass.

\subsection{stellar evolution code}
We employ a MESAbinary module of the Modules for Experiments in Stellar Astrophysics code \citep[MESA, version r-12115,][]{paxt11,paxt13,paxt15,paxt18,paxt19} to calculate the formation and evolution of the source 2A 1822-371. The progenitor of the source is assumed to be a binary system consisting of a NS and a low-mass main-sequence (MS) star in a circular orbit. The code only models the nuclear synthesis and evolution of the MS companion star, and the NS is considered a point mass. The initial chemical composition of the MS companion star is taken to be a solar composition, i.e. $X=0.70, Y=0.28$, and $Z=0.02$.

\begin{table}
\centering

\caption{Some observed data of the source 2A 1822-371.}

\label{teff-filter} \centering
\begin{tabular}{@{}ccc@{}}
  \hline\noalign{\smallskip}
2A 1822-371 & observations & References \\
 \hline\noalign{\smallskip}
orbital period & 5.57 hours &  1,2  \\
donor-star mass & $0.44-0.56~M_{\odot}$& 3  \\
NS mass & $1.61-2.32~M_{\odot}$& 3  \\
orbital-period derivative &$(1.51\pm0.05)\times10^{-10}~\rm s\, s^{-1}$ & 4 \\
mass-accretion rate & $(1.6-5.0)\times10^{-8}~ M_{\odot}\,\rm yr^{-1}$  & 5 \\
 \noalign{\smallskip}\hline
\end{tabular}
References. (1) \cite{maso80}, (2) \cite{whit81}, (3) \cite{muno05}, (4) \cite{anit21}, (5) \cite{van19b}.
\end{table}

During the mass transfer, the mass-growth rate of the NS is limited by Eddington accretion rate ($\dot{M}_{\rm edd}=1.5\times10^{-8}~ M_{\odot}~\rm yr^{-1}$), i.e. the mass-growth rate of the NS $\dot{M}_{\rm ns}={\rm min}(\dot{M}_{\rm edd}, \dot{M_{\rm tr}})$, where $\dot{M_{\rm tr}}$ is the mass-transfer rate. During the accretion, the excess material in unit time ($\dot{M_{\rm tr}}-\dot{M}_{\rm ns}$) is thought to be re-emitted as an isotropic fast wind, carrying away the specific orbital angular momentum of the NS \citep{taur23}.

Using the MESA code, we model the evolution of a LMXB consisting of a NS with a mass of $1.6~M_{\rm \odot}$, and a MS donor star with a mass of $1.0~M_{\rm \odot}$ for two independent models: the CARB MB and the CB disk models. In the CARB MB model, angular momentum loss via gravitational radiation and MB (see also section 2.4) is included. For the CB-disk model, a tiny fraction of the transferred material is assumed to form a CB disk with a constant mass once the mass transfer occurs. Thus the torque originating from the resonant interaction between the CB disk and the binary can extract orbital angular momentum from the orbital motion at a rate of $\dot{J}_{\rm cb}$ (see also equation 9)\footnote{Our inlists and MESA extra files are available at doi: 10.5281/zenodo.8056377}. Furthermore, the gravitation radiation and the standard MB mechanism proposed by \cite{rapp83} with $\gamma=4$ are also considered. The MB mechanism (the CARB MB or the standard MB) would work if the donor star possesses both a convective envelope and a radiative core.

In the calculation, we alter the initial orbital period and/or the CB-disk mass to obtain a model that can match the observed properties of 2A 1822-371, as listed in Table 1. Adopting the CARB MB description, a NS binary with a donor star of $1.75~M_{\rm \odot}$ and an initial orbital period of 0.5 days can evolve into 2A 1822-371 \citep{van21}. Therefore, the initial orbital period ranges from 0.31 (the donor star has already filled its Roche lobe at the evolutionary beginning if the initial orbital period is less than 0.31 days) to 2.0 days in our models. Similar to the inferred CB-disk mass of three black-hole LMXBs \citep{chen19}, the CB-disk mass is assumed to be in the range of $\sim10^{-9}-10^{-7}~M_{\rm \odot}$.

\begin{figure}
	\centering
	\includegraphics[scale=0.34,trim={20 30 0 0},angle=0]{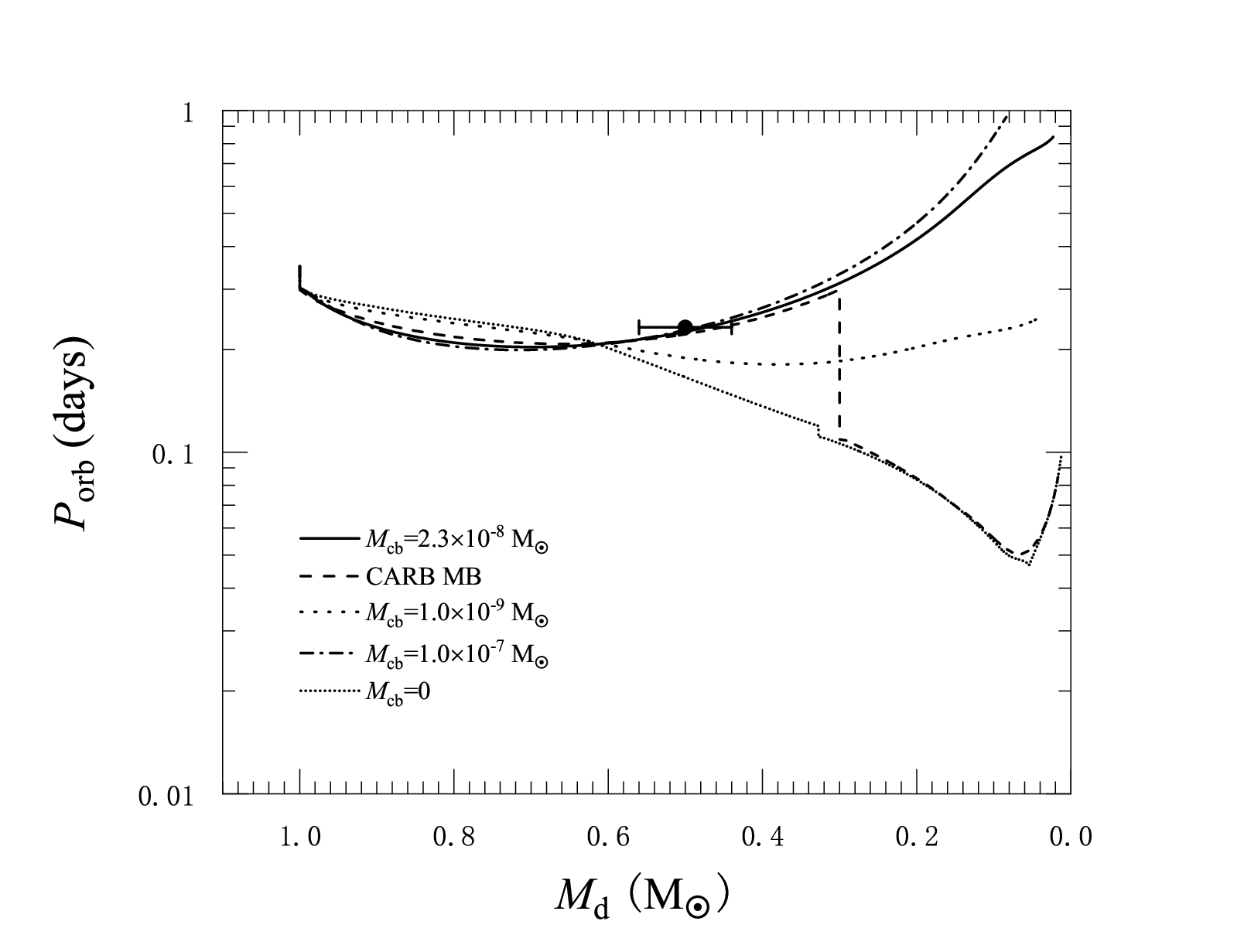}
	\caption{Same as in Figure 2, but for the orbital period vs. donor-star mass diagram. The solid circle represents the observed data of the source 2A 1822-371. }
	\label{fig:sumAH}
\end{figure}

\begin{figure}
	\centering
	\includegraphics[scale=0.34,trim={20 30 0 0},angle=0]{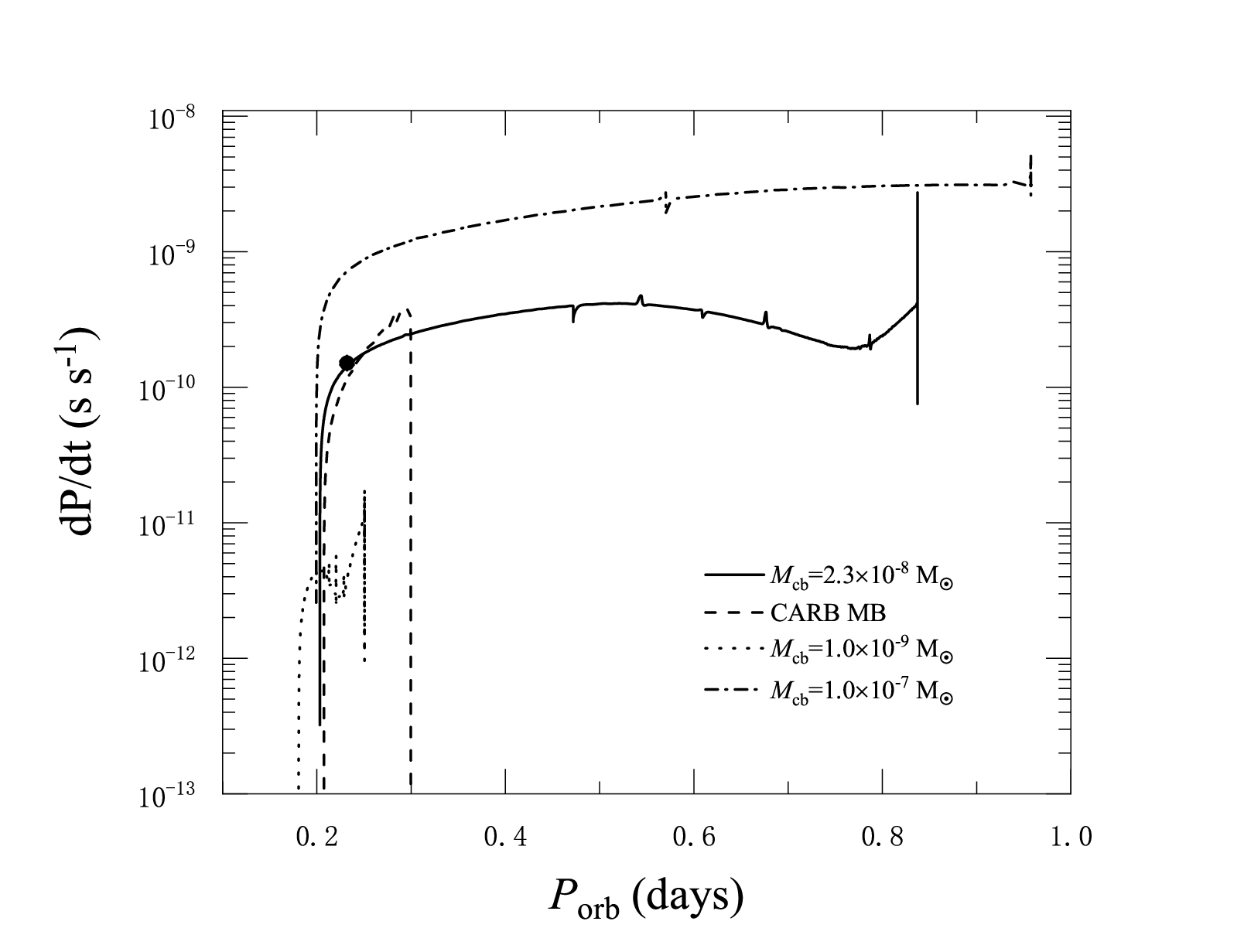}
	\caption{Same as in Figure 2, but for the orbital-period derivative vs. orbital-period diagram. The solid circle represents the observed data of the source 2A 1822-371. }
	\label{fig:sumAH}
\end{figure}

\begin{figure}
	\centering
	\includegraphics[scale=0.34,trim={20 30 0 0},angle=0]{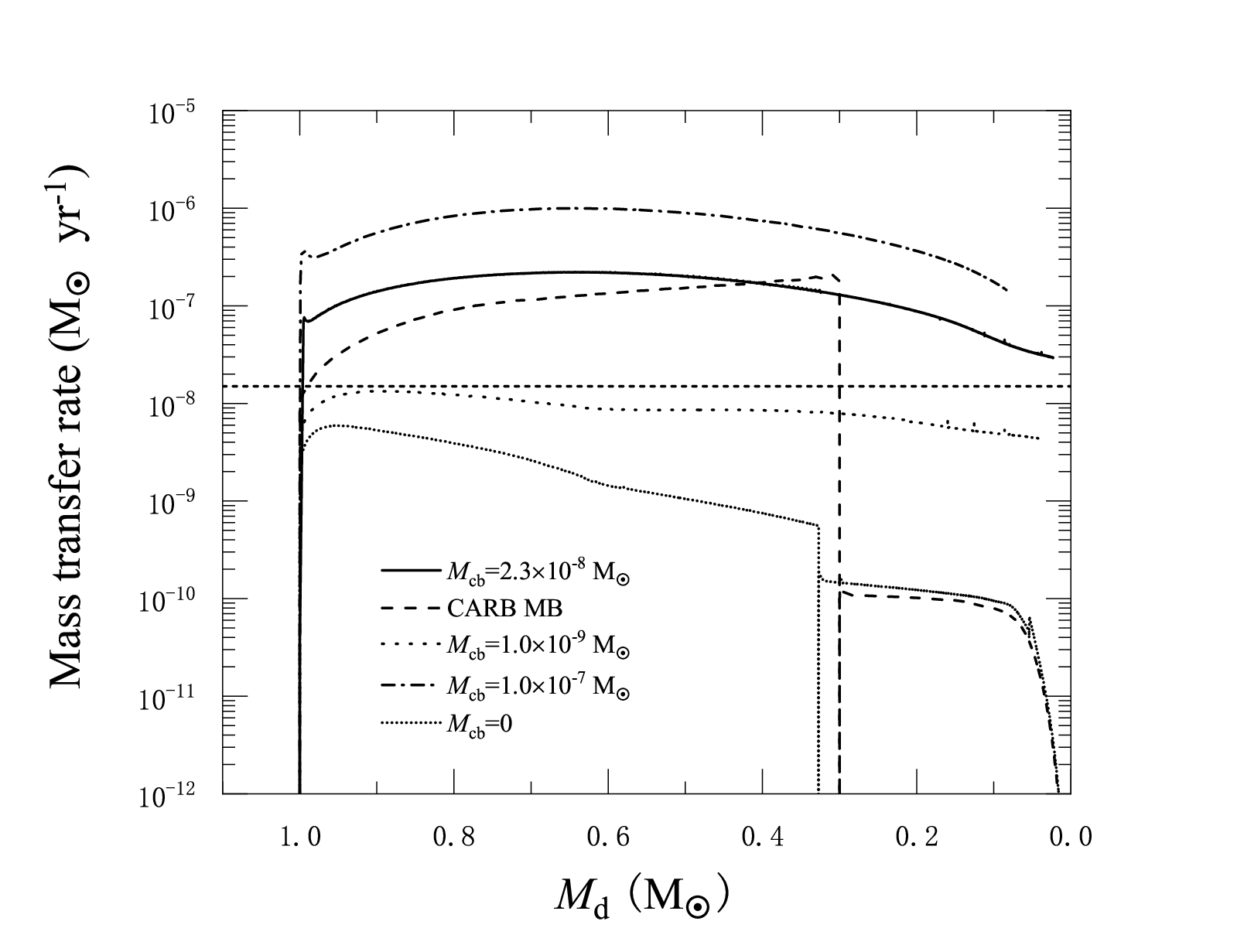}
	\caption{Same as in Figure 2, but for the evolution of mass-transfer rates as a function of the donor-star masses. The horizontal short dashed line denotes the Eddington accretion rate. }
	\label{fig:sumAH}
\end{figure}

\subsection{Simulated results}

Our simulations find that the evolution of a binary consisting of a $1.6~M_{\rm \odot}$ NS and a $1.0~M_{\rm \odot}$ donor star in an initial orbit of 0.35 days can match the observed properties of 2A 1822-371. The evolution of the orbital period and donor-star mass with the stellar age is shown in the upper and bottom panels of Figure 2, respectively. Before the donor star fills its Roche lobe, MB drives the orbit to shrink, and the orbital period decreases to be $\sim0.3$ days. The evolutionary tracks with four different CB-disk masses are the same in this stage. When the stellar age is $t=1.28\times10^7~\rm yr$, the donor star fills its Roche lobe and initiates a mass transfer. Because of a tiny outflow from the mass transfer, a CB disk surrounding the binary forms. Different CB-disk masses result in different angular-momentum-loss rates, driving the binary to evolve along different evolutionary tracks. A heavy CB disk results in a rapid angular-momentum loss and propels the system to evolve into a minimum orbital period in a short timescale. Our simulations indicate that a CB disk with $M_{\rm cb}=2.3\times10^{-8}~M_{\odot}$ can successfully account for the observed orbital period and donor-star mass of 2A 1822-371 when $t=1.59\times10^7~\rm yr$. It seems that the models with $M_{\rm cb}=1.0\times10^{-7}$, and $1.0\times10^{-9}~ M_{\odot}$ also reproduce the observed orbital period in the orbital-expansion stage, however, the corresponding donor-star mass for $M_{\rm cb}=1.0\times10^{-9}$ are much smaller than the observation (see also Figure 3). It is worth emphasizing that the orbital period with $M_{\rm cb}=0$ (i.e., the mass transfer is only driven by the standard MB mechanism) continuously decreases without experiencing an orbital-expansion stage in the observed donor-star mass of 2A 1822-371. At the stellar age of $t=1.1\times10^7~\rm yr$, the CARB MB model can also successfully produce the observed orbital period and donor-star mass of 2A 1822-371. At the current orbital period, the calculated NS masses by the CB-disk and CARB MB models are $1.65~ M_{\odot}$ and $1.71~ M_{\odot}$, respectively.

Figure 3 plots the evolution of LMXBs in the orbital period versus the donor-star mass diagram. Three evolutionary tracks when $M_{\rm cb}=2.3\times10^{-8}~ M_{\odot}$,  $M_{\rm cb}=1.0\times10^{-7}~ M_{\odot}$, and the CARB MB are in agreement with the observed data of 2A 1822-371, while the orbit of the binary continuously lessens for $M_{\rm cb}=1.0\times10^{-9}~ M_{\odot}$, and $M_{\rm cb}=0$, without experiencing an orbital-expansion stage in observed donor-star-mass range.

Figure 4 illustrates the evolution of the orbital-period derivative of LMXBs with three different CB-disk masses and the CARB MB model in the $\dot{P}_{\rm orb}-P_{\rm orb}$ diagram. To compare with the observation, we only plot the evolution of $\dot{P}_{\rm orb}$ in the orbital-expansion stage (the evolutionary track with $M_{\rm cb}=0$ is not included because the orbital-period derivative is negative at the current orbital period). Four orbital periods steadily climb after the minimum orbital periods, resulting in positive orbital-period derivatives. Two orbital-period derivatives predicted by the CB-disk model with $M_{\rm cb}=2.3\times10^{-8}~M_{\odot}$ and the CARB MB model are approximately consistent with the observed value at the current orbital period, while the simulated $\dot{P}_{\rm orb}$ are greater and smaller than the observation for $M_{\rm cb}=1.0\times10^{-7}~M_{\odot}$ and $1.0\times10^{-9}~M_{\odot}$, respectively. Meanwhile, these orbital-period derivatives also emerge an increasing tendency. The factors leading to an increasing $\dot{P}_{\rm orb}$ are very complicated, while the main two factors are the increasing orbital period and the decreasing donor-star mass. Ignoring the contribution of angular-momentum loss, equation (2) yielding $\dot{P}_{\rm orb}\propto P_{\rm orb}/M_{\rm d}$. Therefore, the increasing orbital period and the decreasing donor-star mass should be responsible for the increasing $\dot{P}_{\rm orb}$.

The evolution of the mass-transfer rates of LMXBs is presented in Figure 5. Once the mass transfer starts, the five evolutionary tracks are different because of different angular-momentum-loss rates. It is clear that three models with $M_{\rm cb}=2.3\times10^{-8}~ M_{\odot}$,  $M_{\rm cb}=1.0\times10^{-7}~ M_{\odot}$, and the CARB MB are experiencing a super-Eddington mass transfer in the current donor-star-mass range. A high CB-disk mass of $1.0\times10^{-7}~ M_{\odot}$ produces a high mass-transfer rate of $\sim10^{-7}-10^{-6}~ M_{\odot}\,\rm yr^{-1}$. Therefore, CB disks can accelerate the mass exchange and shorten the evolutionary timescale of LMXBs. The mass-transfer rates produced by the CB-disk model with $M_{\rm cb}=2.3\times10^{-8}~ M_{\odot}$ and the CARB MB model are $1.9\times10^{-7}~M_{\odot}\,\rm yr^{-1}$, and $1.6\times10^{-7}~M_{\odot}\,\rm yr^{-1}$ at the current donor-star mass, respectively. These two mass-transfer rates are approximately one order of magnitude higher than the analytical estimation ($(4.3-5.8)\times10^{-8}~ M_{\odot}\,\rm yr^{-1}$) in Section 2. This difference is because the torque exerted by the CB disk or the CARB MB causes a significant negative orbital-period derivative. In other words, an efficient angular-momentum loss is worthy of a rapier in influencing the orbital evolution of LMXBs as follows: first, it produces a negative $\dot{P}_{\rm orb,aml}$ due to a rapid angular-momentum loss; second, it also results in a positive $\dot{P}_{\rm orb,mt}$ because of a rapid mass transfer from the less massive donor star to the more massive NS (see also equation 2). To compensate for the influence of the negative $\dot{P}_{\rm orb,aml}$, it requires a more high mass-transfer rate to produce the observed $\dot{P}_{\rm orb}$. The simulated mass-transfer rates when $M_{\rm cb}=0$, and $1.0\times10^{-9}~ M_{\odot}$ are always smaller than the Eddington accretion rate, resulting in a negative orbital-period derivative in the observed donor-star-mass range. The main reason is that $\dot{P}_{\rm orb,mt}$ is smaller than $|\dot{P}_{\rm orb,aml}|$, thus the total period derivative ($\dot{P}_{\rm orb}=\dot{P}_{\rm orb,aml}+\dot{P}_{\rm orb,mt}$) is still negative. As $M_{\rm d}\sim0.3~ M_{\odot}$, the mass-transfer rates produced by the CARB and the standard MB models sharply decrease, which arises from the cut-off of MB when the donor stars become fully convective.

\begin{figure}
	\centering
	\includegraphics[scale=0.34,trim={20 30 0 0},angle=0]{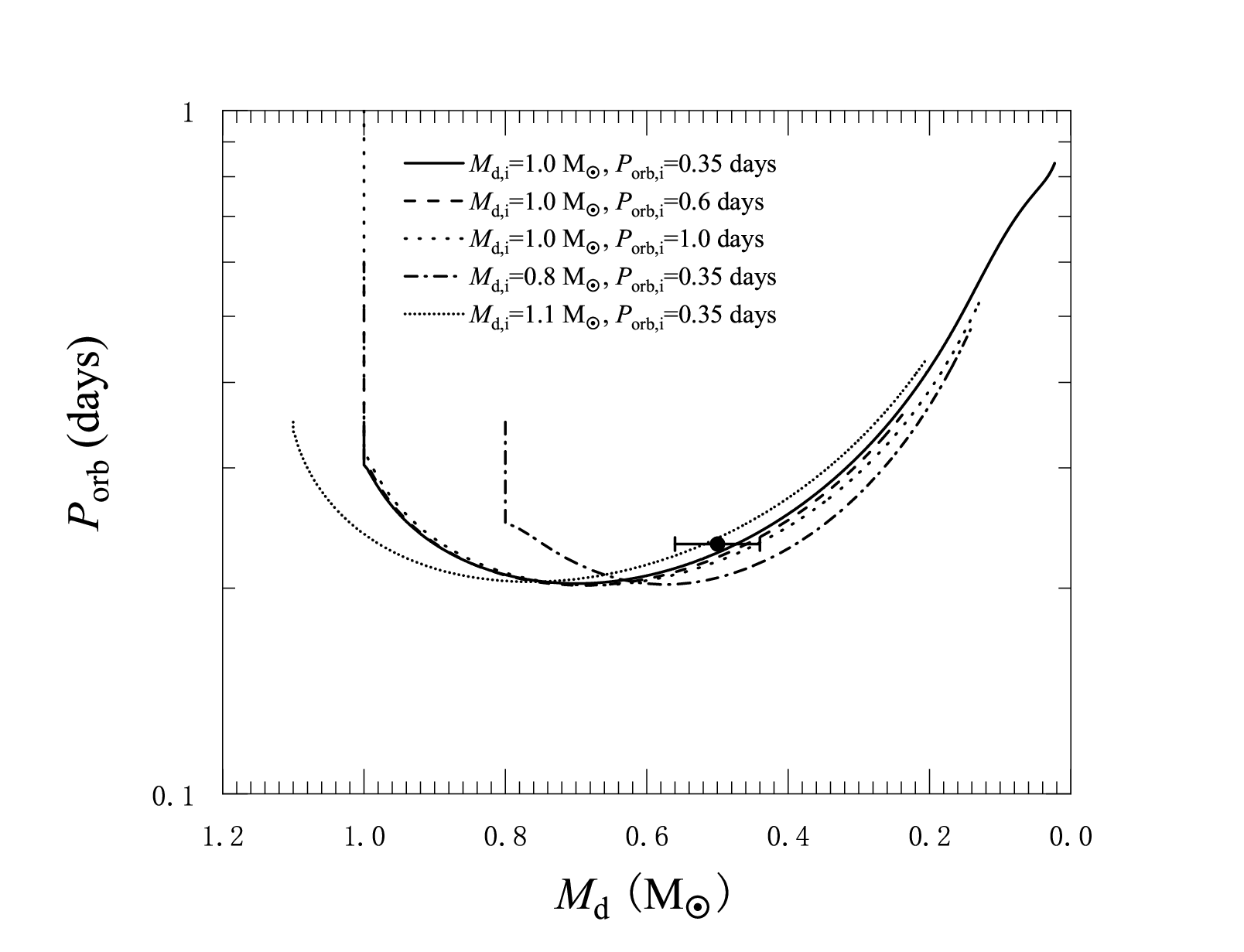}
	\caption{Evolutionary tracks of LMXBs with different initial donor-star masses and initial orbital periods in the orbital period vs. donor-star mass diagram. The solid curves represents our standard model with $M_{\rm d,i}=1.0~M_{\odot}$ and $P_{\rm orb,i}=0.35~\rm days$. A CB disk with a mass of $M_{\rm cb}=2.3\times10^{-8}~M_{\odot}$ is included in all models. The solid circle represents the observed data of the source 2A 1822-371. }
	\label{fig:smAH}
\end{figure}

\begin{figure}
	\centering
	\includegraphics[scale=0.34,trim={20 30 0 0},angle=0]{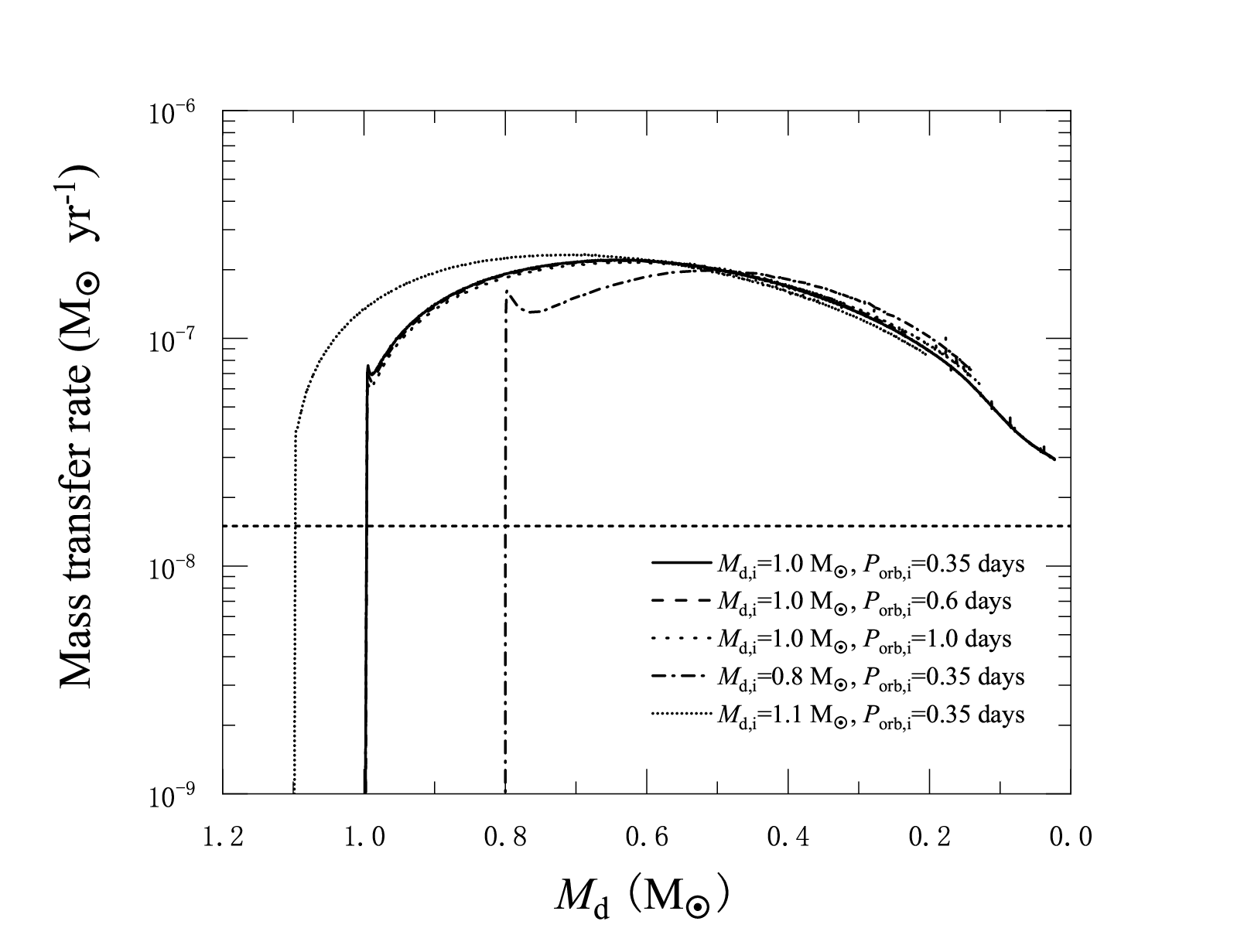}
	\caption{Same as in Figure 6, but for the evolution of mass-transfer rates as a function of the donor-star masses. The horizontal short dashed line denotes the Eddington accretion rate. }
	\label{fig:sumAH}
\end{figure}

\begin{figure}
	\centering
	\includegraphics[scale=0.34,trim={20 30 0 0},angle=0]{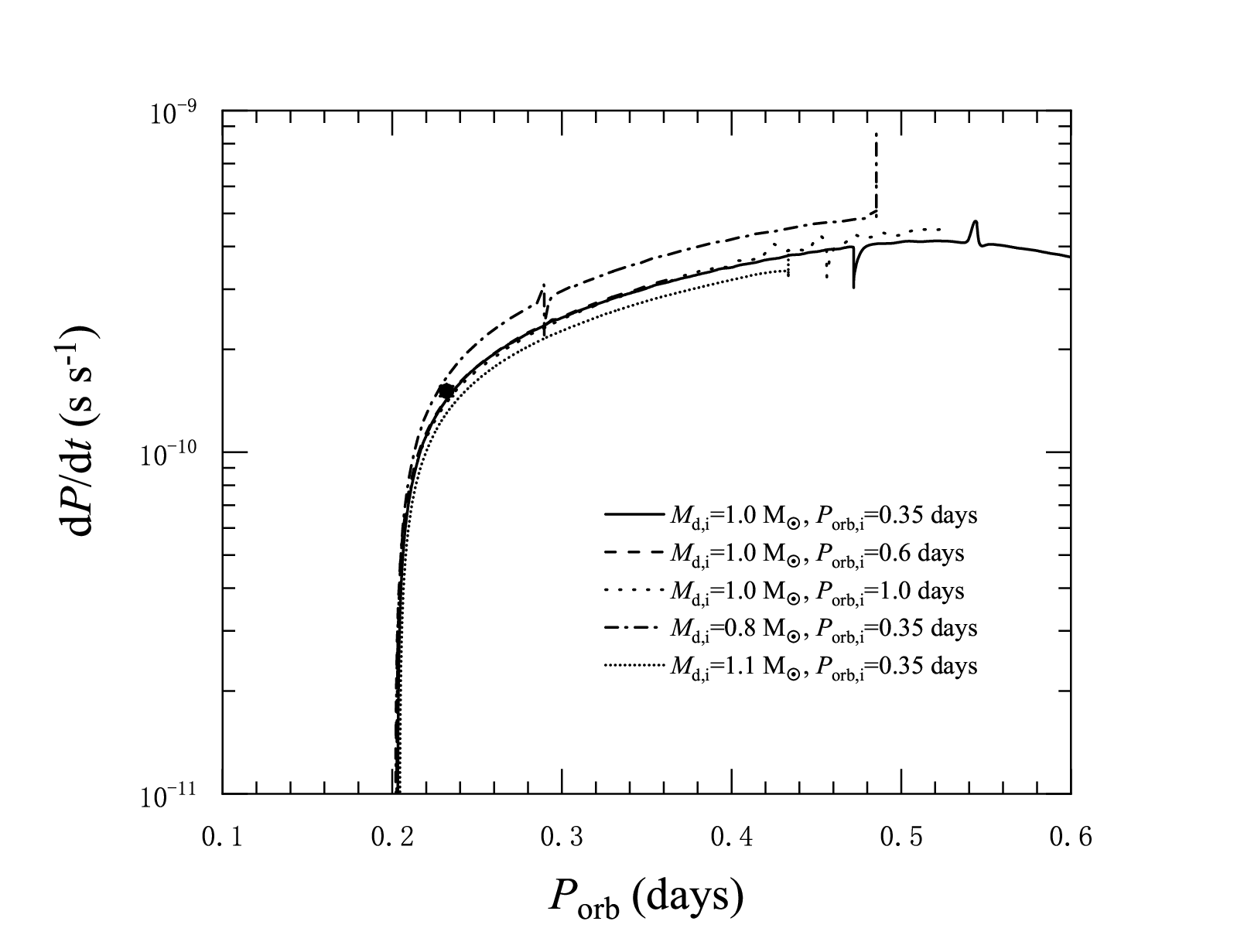}
	\caption{Evolution of orbital period derivatives of LMXBs plotted in Figure 6 in the $\dot{P}_{\rm orb}-P_{\rm orb}$ diagram. For simplicity, we only plot the evolutionary tracks in the orbit-expansion stage. The solid circle represents the observed data of the source 2A 1822-371.}
	\label{fig:sumAH}
\end{figure}


\section{Discussion}
Our standard model proposes that the progenitor of 2A 1822-371 consists of a $1.6~M_{\rm \odot}$ NS and a $1.0~M_{\rm \odot}$ MS companion star in an orbit of 0.35 days, which is surrounded by a CB disk with $M_{\rm cb}=2.3\times10^{-8}~M_{\rm \odot}$ during the mass transfer. In this section, we investigate the influence of initial parameters, including initial donor-star mass ($M_{\rm d,i}$) and initial orbital period ($P_{\rm orb,i}$). Figure 6 depicts the evolution of NS-MS star binaries with different initial companion-star masses and initial orbital periods in the $P_{\rm orb}$-$M_{\rm d}$ diagram. All five models can evolve to the current orbital period in the orbital-expansion stage. It is insensitive to the initial orbital periods whether the binaries can evolve toward the source 2A 1822-371 in the orbital-expansion stage. The models with $(M_{\rm d,i},P_{\rm orb,i})=(1.0~M_{\rm \odot},0.6~\rm days)$, and $(1.0~M_{\rm \odot},1.0~\rm day)$ can also reproduce the observed orbital period and donor-star mass. A system with a relatively high donor-star mass of $1.1~M_{\rm \odot}$ and the same orbital period of 0.35 days is also the potential progenitor of 2A 1822-371. Three models with $(M_{\rm d,i},P_{\rm orb,i})=(1.0~M_{\rm \odot},0.35~\rm days)$, $(1.0~M_{\rm \odot},0.6~\rm days)$, and $(1.0~M_{\rm \odot},1.0~\rm day)$ have similar evolutionary tendency and slope, resulting in a similar $\dot{P}_{\rm orb}$ (see also Figure 8).

To account for the current NS mass of 2A 1822-371, both the CB-disk model and the CARB MB model found that the NS was born massive ($\sim1.6~M_{\rm \odot}$). This is very similar to PSR J1614-2230 and PSR J1640+2224, which were proposed to be born with masses of $\sim1.95~M_{\rm \odot}$ \citep{taur11} and $>2.0~M_{\rm \odot}$ \citep{deng20}, respectively. Because of a relatively low mass-transfer rate in the early stage (see also Figure 5), the accretion efficiency of the CARB MB model is slightly higher than that of the CB-disk model.

The similar evolutionary tendencies of the five curves in Figure 6 should originate from similar mass-transfer rates. Figure 7 summarizes the evolution of mass-transfer rates as a function of the donor-star masses. It seems that the model with $(M_{\rm d,i},P_{\rm orb,i})=(0.8~M_{\rm \odot}, 0.35~\rm days)$ possesses a maximum mass-transfer rate at the current stage of 2A 1822-371. In the observed donor-star-mass range, five models have similar mass-transfer rates ($1.7-2.1\times10^{-7}~M_{\rm \odot}~\rm yr^{-1}$), which are higher than theoretical estimation in section 2. A high mass-transfer rate naturally results in rapid orbital expansion, which can vanquish the orbital shrinkage originating from angular-momentum loss. Therefore, the orbital evolution of five models emerges an expansion tendency.

Figure 8 shows the evolution of LMXBs with different initial donor-star masses and initial orbital periods in the $\dot{P}_{\rm orb}-P_{\rm orb}$ diagram. It seems that a small donor-star mass seems produces a high orbital-period derivative at the current orbital period of 2A 1822-371. Three models with a $1.0~M_{\rm \odot}$ donor star and initial orbital periods of $0.35, 0.6$, and $1.0~\rm day$ can approximately reproduce the observed orbital-period derivative of 2A 1822-371. In a word, the peculiar observed properties of this source are probably associated with three initial parameters, including the initial donor-star mass, the initial orbital period, and the CB-disk mass.

To produce the observed orbital-period derivative of 2A 1822-371, our CB-disk model predicts a relatively high mass-transfer rate of $1.9\times10^{-7}~M_{\rm \odot}~\rm yr^{-1}$. Such a mass-transfer rate is higher than the mass-accretion rate ($1.6-5.0\times10^{-8}~ M_{\odot}\,\rm yr^{-1}$) derived from the intrinsic luminosity of 2A 1822-371 \citep{van19b}. This implies a much outflow during the mass transfer, which would absorb a large fraction of X-ray radiation, resulting in a low observed luminosity. Meanwhile, because the orbital inclination angle of this source is in the range of $81^{\circ}-85^{\circ}$ \citep{hein01,ji11}, an edge-on accretion disk may be responsible for the low X-ray luminosity in observations.

There exist some uncertainties in our simulation. The orbital-period derivative produced by the mass transfer in our standard model is $\dot{P}_{\rm orb,mt}=-3P_{\rm orb}\frac{\dot{M}_{\rm d}}{M_{\rm d}}[1-q\beta-\frac{q(1-\beta)}{3(1+q)}]\approx6.5\times10^{-10}~\rm s\,s^{-1}$ when we take $M_{\rm d}=0.5~M_{\odot}$, $M_{\rm NS}=1.6~M_{\odot}$, $\beta=0.1$, and $\dot{M}_{\rm d}=1.9\times10^{-7}~M_{\rm \odot}~\rm yr^{-1}$. Because the observed donor-star mass $M_{\rm d}=0.5\pm0.06~M_{\odot}$, it has an uncertainty of $12\%$. The observed mass-transfer rate is $\dot{M}_{\rm d}=(3.3\pm1.7)\times10^{-8}~M_{\rm \odot}~\rm yr^{-1}$, which has an approximate uncertainty of $50\%$. We can approximately estimate the minimum uncertainty of $\dot{P}_{\rm orb,mt}$ as $\sqrt{(12\%)^{2}+(50\%)^{2}}=51\%$ \footnote{Here, the uncertainty of $1-q\beta-\frac{q(1-\beta)}{3(1+q)}$ is not considered. We also ignore the uncertainty of angular-momentum loss by MB because $|\dot{J}_{\rm mb}/J|$ is two orders of magnitude smaller than $\dot{P}_{\rm orb}/P$ (see also section 2.1).}, thus $\dot{P}_{\rm orb,mt}\approx(6.5\pm3.3)\times10^{-10}~\rm s\,s^{-1}$. The orbital-period derivative produced by the CB disk is \citep{chen19}
\begin{equation}
\dot{P}_{\rm orb,cb}=\frac{3\dot{J}_{\rm cb}P_{\rm orb}}{J}=-6\pi\alpha\frac{M_{\rm cb}(M_{\rm d}+M_{\rm NS})}{M_{\rm d}M_{\rm NS}}\left(\frac{H}{R}\right)^{2}\frac{a}{R}.
\end{equation}
Taking $M_{\rm cb}=2.3\times 10^{-8}~M_{\odot}$, we have $\dot{P}_{\rm orb,cb}=-4.95\times10^{-10}~\rm s\,s^{-1}$. There also exist some uncertainties in the CB-disk mass, and the disk parameters $\alpha(H/R)^{2}a/R$. Black-hole LMXB XTE J1118+480 was detected a rapid orbital decay with an orbital-period derivative as $\dot{P}_{\rm orb}=-(6.01\pm1.81)\times10^{-11}~\rm s\,s^{-1}$ \citep{gonz14}, and its surrounding CB-disk mass was constrained to be $10^{-9}~M_{\odot}$. Taking $H/R=0.1$, $a/R=1/2.3$, $M_{\rm d}=0.18~M_{\odot}$, and the black-hole mass $M_{\rm BH}=8~M_{\odot}$, it yields $\alpha=0.13\pm0.04$ according to equation (10). Therefore, $\alpha=0.1$ that we adopt has at least an uncertainty of $\sim30\%$. Those uncertainties of other parameters are difficult to estimate. For simplicity, the minimum uncertainty of $\dot{P}_{\rm orb,cb}$ is $\sim30\%$, i.e.$\dot{P}_{\rm orb,cb}=(-4.95\pm1.49)\times10^{-10}~\rm s\,s^{-1}$. Because $\dot{P}_{\rm orb}=\dot{P}_{\rm orb,mt}+\dot{P}_{\rm orb,cb}$ , the minimum error of $\dot{P}_{\rm orb}$ is $\delta\dot{P}_{\rm orb}=\sqrt{(\delta\dot{P}_{\rm orb,mt})^{2}+(\delta\dot{P}_{\rm orb,cb})^{2}}\approx3.6\times10^{-10}~\rm s\,s^{-1}$. As a consequence, the orbital-period derivative is $\dot{P}_{\rm orb}=(1.55\pm3.6)\times10^{-10}~\rm s\,s^{-1}$, in which the expected error exceeds our simulated value by at least a factor of 2.

Compared with the work of \cite{xing19}, NS-MS star binary with a CB disk can evolve into 2A 1822-371 in a wide initial orbital-period range ($0.35-1.0$ days). Since the angular-momentum-loss rate due to MB by coupling between the strong magnetic field and an irradiation-driving wind is smaller than that by a CB disk, it requires a relatively short initial orbital period to reproduce the observed orbital period and orbital-period derivative. As a consequence, the excited wind-driving mass-transfer model found that the progenitor of 2A 1822-371 is a narrow orbit system with a short orbital period of 0.4 days when the surface magnetic field of the donor star is 900 G \citep{xing19}. Another distinction between these two models is the evolutionary tendency of the mass-transfer rates. In the observed donor-star-mass range, the CB disk model predicts a decreasing mass-transfer rate (see also Figures 5 and 7); however, the one obtained by the excited wind-driving mass-transfer model is increasing \citep{xing19}. Since the mass transfers of both models are super-Eddington, it is hard to test the validity of these two models according to the observed X-ray luminosities.

If a CB disk exists around source 2A 1822-371, the infrared radiation might confirm its existence, like GG Tau \citep{rodd96}. In particular, recent works performed by the Wide-Field Infrared Survey Explorer confirmed that three black-hole LMXBs XTE J1118+480, A0620-00, GRS 1915+105, and NS LMXB 3A 1728-247 are surrounded by CB disks \citep{wang14}. However, the orbital period (1160.8 days) of NS LMXB 3A 1728-247 is much longer than that of 2A 1822-371. Therefore, it still has no observable evidence for the existence of such a CB disk around NS LMXBs similar to 2A 1822-371. Furthermore, the orbital inclination angle of 2A 1822-371 is in the range of $81^{\circ}-85^{\circ}$ \citep{hein01,ji11}, it is challenging to detect an edge-on CB disk. We expect that the powerful new infrared capabilities of JWST will confirm whether or not a CB disk encloses the source 2A 1822-371.

In theory, a CB disk around LMXBs can not only cause a rapid orbital expansion but also lead to a fast orbital shrinkage. According to equation (10), the angular-momentum loss by a CB disk results in a negative orbital-period derivative as
\begin{equation}
\begin{aligned}
\dot{P}_{\rm orb,cb}=-2.2\times10^{-11}\left(\frac{M_{\rm cb}}{10^{-9}~M_{\odot}}\right)
\left(\frac{0.38~M_{\odot}}{\mu}\right)\\
\left(\frac{\alpha(H/R)^{2}}{10^{-3}}\right)\left(\frac{2.3}{R/a}\right)~\rm s\,s^{-1},
\end{aligned}
\end{equation}
where $\mu=M_{\rm d}M_{\rm NS}/(M_{\rm d}+M_{\rm NS})$ is the reduced mass of LMXBs. Taking $\beta=1$, the mass transfer produces a positive orbital-period derivative as
\begin{equation}
\begin{aligned}
\dot{P}_{\rm orb,mt}=2.6\times10^{-11}\left(\frac{P_{\rm orb}}{5.57~\rm hours}\right)\left(\frac{\dot{M}_{\rm d}}{-10^{-8}~M_{\odot}\,\rm yr^{-1}}\right)\\
\left(\frac{0.5~M_{\odot}}{M_{\rm d}}\right)(1-q)~\rm s\,s^{-1}.
\end{aligned}
\end{equation}
The orbital evolution fates of LMXBs would depend on the competition between $\dot{P}_{\rm orb,cb}$ and $\dot{P}_{\rm orb,mt}$. For a LMXB like 2A 1822-371, $M_{\rm d}=0.5~M_{\odot}$, $\mu=0.38~M_{\odot}$, $P_{\rm orb}=5.57~\rm hours$, we have $\dot{P}_{\rm orb,cb}=-2.2\times10^{-11}~\rm s\,s^{-1}$ if the system is surrounded by a CB disk with a mass similar to XTE J1118+480. Similarly, $\dot{P}_{\rm orb,mt}=(0.26-2.6)\times10^{-11}~\rm s\,s^{-1}$ if the system possesses a mass ratio ($q=0.31$) same to 2A 1822-371 and a mass-transfer rate in the range from $10^{-9}~M_{\odot}\,\rm yr^{-1}$ to $10^{-8}~M_{\odot}\,\rm yr^{-1}$. As a result, the total orbital-period derivative ( $\dot{P}_{\rm orb}=\dot{P}_{\rm orb,cb}+\dot{P}_{\rm orb,mt}$) is in the range from $-1.94\times10^{-11}~\rm s\,s^{-1}$ to $0.4\times10^{-11}~\rm s\,s^{-1}$. Therefore, the LMXB would evolve toward a wide orbit binary like 2A 1822-371 for a high mass-transfer rate. Otherwise, its descendant is a narrow orbit system like black-hole LMXBs XTE J1118+480 and A0620-00 \citep{xu18,chen19}.

\section{Summary}
According to the observed orbital-period derivative of 2A 1822-371, our analysis finds that the mass-transfer rate must be in the range of $(4.3-5.8)\times10^{-8}~ M_{\odot}\,\rm yr^{-1}$, which is non-sensitive to the accretion efficiency of the NS. Besides gravitational radiation and MB, it requires an additional mechanism to efficiently extract angular momentum from the system, resulting in a rapid mass transfer.

In this work, we attempt to investigate whether a surrounding CB disk can explain the anomalous orbital-period derivative of 2A 1822-371. Adopting typical CB-disk parameters as $\alpha=0.1$, $H/R=0.1$, and $R/a=2.3$, detailed stellar evolution models show that a CB disk with a mass of $M_{\rm cb}=2.3\times10^{-8}~M_{\rm \odot}$ can reproduce the observed orbital period, donor-star mass, and orbital-period derivative of 2A 1822-371. However, the expected error of the orbital-period derivative exceeds our simulated value by at least a factor of 2. According to our simulations, the progenitor of the source 2A 1822-371 is probably a binary system including a $1.6~M_{\rm \odot}$ NS and a $1.0~M_{\rm \odot}$ MS companion star in an initial orbit of 0.35 days. Although the angular-momentum loss by the CB disk gives rise to a rapid orbital shrinkage, a high mass-transfer rate of $1.9\times10^{-7}~M_{\rm \odot}~\rm yr^{-1}$ from the less massive donor star to the more massive NS causes a more significant orbital expansion. It is noteworthy that there is no direct observational evidence for 2A 1822-371 to possess such a high mass-transfer rate.

Our calculations also find that CB disks can accelerate the mass exchange of LMXBs, and alter their evolutionary fates, which strongly depend on the CB-disk masses and mass-transfer rate. For a system similar to 2A 1822-371, a CB disk with a mass of $M_{\rm cb}=10^{-9}~M_{\rm \odot}$ would cause its orbit to widen for a high mass-transfer of $10^{-8}~M_{\odot}\,\rm yr^{-1}$. However, the orbit of the system would continuously shrink for the same CB disk and a low mass-transfer rate of $10^{-9}~M_{\odot}\,\rm yr^{-1}$. We have to stress that observable evidence of such a CB disk is still absent in NS LMXBs similar to 2A 1822-371.

It is worth emphasizing that the CARB MB model not only can interpret the observed characteristics of all persistent NS LMXBs \citep{van19a,van21} and binary pulsars \citep{deng21}, but also reproduce the anomalous orbital-period derivative of 2A 1822-371. Therefore, it might be the best MB model to replace the standard MB model in the binary evolution model in the future.

\begin{acknowledgements}
We are extremely grateful to the anonymous referee for very constructive and detailed comments that improved this
manuscript. This work was partly supported by the National Natural Science Foundation of China (under grant numbers 12273014, 12373044, 11733009), Natural Science Foundation (under grant number ZR2021MA013) of Shandong Province.
\end{acknowledgements}

\end{document}